\documentclass[11pt]{article}
\usepackage{graphicx}
\begin{document}

\catcode`@=11
\long\def\@caption#1[#2]#3{\par\addcontentsline{\csname
  ext@#1\endcsname}{#1}{\protect\numberline{\csname
  the#1\endcsname}{\ignorespaces #2}}\begingroup
    \small
    \@parboxrestore
    \@makecaption{\csname fnum@#1\endcsname}{\ignorespaces #3}\par
  \endgroup}
\catcode`@=12
\newcommand{\newc}{\newcommand}
\newc{\lat}{{\ell at}}
\newc{\one}{{\bf 1}}
\newc{\mgut}{M_{\rm GUT}}
\newc{\mzero}{m_0}
\newc{\mhalf}{M_{1/2}}
\newc{\five}{{\bf 5}}
\newc{\fivebar}{{\bf\bar 5}}
\newc{\ten}{{\bf 10}}
\newc{\tenbar}{{\bf\bar{10}}}
\newc{\sixteen}{{\bf 16}}
\newc{\sixteenbar}{{\bf\bar{16}}}
\newc{\gsim}{\lower.7ex\hbox{$\;\stackrel{\textstyle>}{\sim}\;$}}
\newc{\lsim}{\lower.7ex\hbox{$\;\stackrel{\textstyle<}{\sim}\;$}}
\newc{\gev}{\,{\rm GeV}}
\newc{\mev}{\,{\rm MeV}}
\newc{\ev}{\,{\rm eV}}
\newc{\kev}{\,{\rm keV}}
\newc{\tev}{\,{\rm TeV}}
\newc{\mz}{m_Z}
\newc{\mw}{m_W}
\newc{\mpl}{M_{Pl}}
\newc{\mh}{m_h}
\newc{\mA}{m_A}
\newc{\tr}{\mbox{Tr}}
\def\sfrac#1#2{{\textstyle\frac{#1}{#2}}}
\newc{\chifc}{\chi_{{}_{\!F\!C}}}
\newc\order{{\cal O}}
\newc\CO{\order}
\newc\CL{{\cal L}}
\newc\CY{{\cal Y}}
\newc\CH{{\cal H}}
\newc\CM{{\cal M}}
\newc\CF{{\cal F}}
\newc\CD{{\cal D}}
\newc\CN{{\cal N}}
\newc{\eps}{\epsilon}
\newc{\re}{\mbox{Re}\,}
\newc{\im}{\mbox{Im}\,}
\newc{\invpb}{\,\mbox{pb}^{-1}}
\newc{\invfb}{\,\mbox{fb}^{-1}}
\newc{\yddiag}{{\bf D}}
\newc{\yddiagd}{{\bf D^\dagger}}
\newc{\yudiag}{{\bf U}}
\newc{\yudiagd}{{\bf U^\dagger}}
\newc{\yd}{{\bf Y_D}}
\newc{\ydd}{{\bf Y_D^\dagger}}
\newc{\yu}{{\bf Y_U}}
\newc{\yud}{{\bf Y_U^\dagger}}
\newc{\ckm}{{\bf V}}
\newc{\ckmd}{{\bf V^\dagger}}
\newc{\ckmz}{{\bf V^0}}
\newc{\ckmzd}{{\bf V^{0\dagger}}}
\newc{\X}{{\bf X}}
\newc{\bbbar}{B^0-\bar B^0}
\def\bra#1{\left\langle #1 \right|}
\def\ket#1{\left| #1 \right\rangle}
\newc{\sgn}{\mbox{sgn}\,}
\newc{\m}{{\bf m}}
\newc{\msusy}{M_{\rm SUSY}}
\newc{\munif}{M_{\rm unif}}
\newc{\slepton}{{\tilde\ell}}
\newc{\Slepton}{{\tilde L}}
\newc{\sneutrino}{{\tilde\nu}}
\newc{\selectron}{{\tilde e}}
\newc{\stau}{{\tilde\tau}}
\newc{\vbb}{\beta\beta 0\nu}
\newc{\vckm}{V_{\!\mbox{\tiny CKM}}}
\newc{\mbbeff}{m_{\beta\beta}^{\mbox{\tiny eff}}}
%
%
\def\NPB#1#2#3{Nucl. Phys. {\bf B#1} (19#2) #3}
\def\PLB#1#2#3{Phys. Lett. {\bf B#1} (19#2) #3}
\def\PLBold#1#2#3{Phys. Lett. {\bf#1B} (19#2) #3}
\def\PRD#1#2#3{Phys. Rev. {\bf D#1} (19#2) #3}
\def\PRL#1#2#3{Phys. Rev. Lett. {\bf#1} (19#2) #3}
\def\PRT#1#2#3{Phys. Rep. {\bf#1} (19#2) #3}
\def\ARAA#1#2#3{Ann. Rev. Astron. Astrophys. {\bf#1} (19#2) #3}
\def\ARNP#1#2#3{Ann. Rev. Nucl. Part. Sci. {\bf#1} (19#2) #3}
\def\MPL#1#2#3{Mod. Phys. Lett. {\bf #1} (19#2) #3}
\def\ZPC#1#2#3{Zeit. f\"ur Physik {\bf C#1} (19#2) #3}
\def\APJ#1#2#3{Ap. J. {\bf #1} (19#2) #3}
\def\AP#1#2#3{{Ann. Phys. } {\bf #1} (19#2) #3}
\def\RMP#1#2#3{{Rev. Mod. Phys. } {\bf #1} (19#2) #3}
\def\CMP#1#2#3{{Comm. Math. Phys. } {\bf #1} (19#2) #3}
\relax
%
%
%
\def\beq{\begin{equation}}
\def\eeq{\end{equation}}
\def\bea{\begin{eqnarray}}
\def\eea{\end{eqnarray}}
%
%
%
\newc{\ie}{{\it i.e.}}          \newc{\etal}{{\it et al.}}
\newc{\eg}{{\it e.g.}}          \newc{\etc}{{\it etc.}}
\newc{\cf}{{\it c.f.}}
\def\smuon{{\tilde\mu}}
\def\neut{{\tilde N}}
\def\char{{\tilde C}}
\def\bino{{\tilde B}}
\def\wino{{\tilde W}}
\def\higgsino{{\tilde H}}
\def\sneut{{\tilde\nu}}
%
%
%
%
\def\slash#1{\rlap{$#1$}/} 
\def\Dsl{\,\raise.15ex\hbox{/}\mkern-13.5mu D} 
\def\delsl{\raise.15ex\hbox{/}\kern-.57em\partial}
\def\Ksl{\hbox{/\kern-.6000em\rm K}}
\def\Asl{\hbox{/\kern-.6500em \rm A}}
\def\Qsl{\hbox{/\kern-.6000em\rm Q}}
\def\gradsl{\hbox{/\kern-.6500em$\nabla$}}
%
%
%
\def\bar#1{\overline{#1}}
\def\vev#1{\left\langle #1 \right\rangle}
\def\mueff{\mu_{\rm eff}}
\def\Bmueff{B_{\mu,\rm eff}}
%

\begin{titlepage}
\begin{flushright}

\end{flushright}
\vskip 2cm
\begin{center}
{\large\bf
The SUSY Higgs Mass: the Singlet Saves the Day}
\vskip 1cm
{\normalsize\bf
Antonio Delgado\footnote[1]{antonio.delgado@nd.edu}}, {\bf Christopher Kolda\footnote[2]{ckolda@nd.edu}}, {\bf J. Pocahontas Olson\footnote[3]{jspeare@nd.edu}}, \\ and {\bf 
Alejandro de la Puente\footnote[4]{adelapue@nd.edu}\\
\vskip 0.5cm
{\it Department of Physics, University of Notre Dame\\
Notre Dame, IN~~46556, USA}\\[0.1truecm]
}

\end{center}
\vskip .5cm

\begin{abstract}
We present a generalization of the Next-to-Minimal Supersymmetric Standard Model (NMSSM), with an explicit $\mu$-term and a supersymmetric mass for the singlet superfield, as a route to alleviating the little hierarchy problem of the Minimal Supersymmetric Standard Model (MSSM). Though this model does not address the $\mu$-problem of the MSSM, we are able to generate masses for the lightest neutral Higgs boson up to $140\gev$ with top squarks below the TeV scale, all couplings perturbative to the gauge unification scale, and with no need to fine tune parameters in the scalar potential. This model, which we call the S-MSSM, more closely resembles the MSSM phenomenologically than the NMSSM as usually defined.

\end{abstract}

\end{titlepage}

\setcounter{footnote}{0}
\setcounter{page}{1}
\setcounter{section}{0}
\setcounter{subsection}{0}
\setcounter{subsubsection}{0}

In the Minimal Supersymmetric Standard Model (MSSM), the simplest implementation of supersymmetry (SUSY) on to the structure of the Standard Model, the Higgs sector is tightly constrained by a combination of experimental searches and theoretical bounds. Though there are two Higgs doublets ($H_u$, $H_d$) and 5 physical degrees of freedom in the Higgs sector of the MSSM, the lightest neutral scalar ($h$) has properties that usually mimic the Higgs boson of the Standard Model, both in its production and decay. However, unlike the Standard Model Higgs, this SUSY Higgs has its mass controlled by gauge couplings, leading to the well-known, tree-level upper bound $m_h\leq m_z\cos 2\beta$, where $\tan\beta$ is the ratio of the vevs of the two Higgs doublets, $v_u/v_d$. It is equally well-known that this bound is lifted by large one-loop radiative corrections, the largest of which come from loops of top quarks and squarks.

Given the lower bound on the Standard Model Higgs mass from the combined LEP analysis, $m_h>114\gev$~\cite{LEP}, it is clear that a Standard Model-like SUSY Higgs must get a considerable fraction of its mass from these radiative corrections. In realistic models, this usually means forcing the top squarks to sit at scales above a TeV. But these same heavy stops generate quadratic corrections to the Higgs potential that force a careful tuning of the mass parameters in order to generate a Standard Model vev of $174\gev$. This tuning has come to be called the ``little hierarchy" problem, an unfortunate development in a model promoted for its ability to solve the (albeit much larger) Standard Model hierarchy problem.

Many attempts have been made to alleviate this predicament, either: by hiding the lightest Higgs from experimental discovery~\cite{gunion}; by lowering the cut-off scale either explicitly~\cite{lowcutoff} or by imposing strong coupling on the theory~\cite{fathiggs}; 
or by extending the MSSM through additional particles, operators~\cite{dst,carena}, or symmetries~\cite{bgk}, all in the service of lifting the Higgs mass. In this latter vein, one popular route, going back several decades, is to add gauge singlets to the MSSM and to raise the light Higgs mass by mixing the singlets with the usual Higgs fields~\cite{quiesp}. The simplest implementation of this approach contains only one singlet and is known as the Next-to-Minimal SUSY Standard Model (NMSSM). In the NMSSM  (see~\cite{nmssmreview} for a good review), one discards the $\mu$-term of the MSSM superpotential in favor of a gauge singlet superfield whose scalar vev provides an effective $\mu$-term:
\begin{equation}
W = W_{\rm Yukawa} + \lambda S H_u H_d + \frac{\kappa}{3} S^3,
\label{eq1}
\end{equation}
where $W_{\rm Yukawa}$ represents the couplings of $H_u$ and $H_d$ to the quark and lepton superfields, and $\lambda$ and $\kappa$ are arbitrary coupling constants. (We use the standard convention that the SU(2) contraction $H_u H_d = H_u^+ H_d^- - H_u^0 H_d^0$.)
Because the singlet vev is generated by soft SUSY-breaking mass terms in the scalar potential, it is naturally of the order of the soft masses and therefore of the electroweak scale, solving the ``$\mu$ problem" of the MSSM. 

Though the NMSSM is a very attractive model in principle, it is constrained by experimental data to live in very particular corners of its parameter space. In minimizing the full scalar potential of the NMSSM, there are two large regions of parameter space which must be avoided: one in which $S$ receives a very small vev, leading to a small $\mu$-term, which is ruled out by the absence of light charginos and neutralinos; the other in which $S$ is driven to large values, over-stabilizing the Higgs potential and preventing electroweak symmetry breaking. A phenomenologically viable model must live between these extremes, while generating Higgs sectors consistent with current experimental constraints. 

Within the NMSSM, there are three primary ways to obtain consistency with the Higgs searches at LEP. The first is to push top squark masses and mixing parameters ($A$-terms) to super-TeV scales, generating large radiative corrections which lift the light Higgs mass; we reject this idea, as it simply moves the little hierarchy problem from the MSSM to the NMSSM. A second proposal~\cite{gunion} is to mix the singlets with the Higgs doublets in such a way as to generate very light Higgs bosons that are difficult to produce and detect at LEP, thereby avoiding all existing Higgs search constraints. The third proposal, which has been studied for many years, is to use the $\lambda$-term to lift the lightest Higgs mass~\cite{quiesp}. It is well known that, at tree level in the NMSSM:
\begin{equation}\label{nmssmresult}
m^2_{h^0} \leq \mz^2\cos^2 2\beta + \frac{\lambda^2 v^2}{2}\sin^2 2\beta
\end{equation}
where $v=174\gev$ and $\tan\beta = \langle H_u\rangle/\langle H_d \rangle$.  This well-known expression allows the lightest Higgs mass to be raised above the MSSM limit, though only at $\tan\beta$ close to one, where the MSSM contribution is in fact minimized. Further, it is bounded by the usual requirement that $\lambda$ remain perturbative all the way to the gauge unification scale so as not to disrupt the gauge coupling unification of the MSSM.

But the superpotential of the NMSSM is not the most general superpotential one encounters when extending the MSSM by the addition of one gauge singlet. In fact, if we only impose gauge symmetries and $R$-parity, the most general, renormalizable superpotential is:
\beq \label{Wfull}
W=W_{\rm Yukawa} + (\mu+\lambda S) H_u H_d + \frac{\mu_S}{2} S^2 + \frac{\kappa}{3} S^3 + \xi S
\eeq
In particular, the most general superpotential contains explicit mass terms both for the $H_uH_d$ pair (the usual $\mu$-term) and for the singlet itself ($\mu_s$), as well as a tadpole term for the singlet. In the analysis that follows, we will allow $\mu$ and $\mu_S$ to take arbitrary values at or below the TeV scale, independent of the vev of $S$, and so we give up any attempt at solving the $\mu$-problem (or $\mu_s$-problem). But in its place we will find a lightest Higgs more easily lifted above the LEP bound. Though there is no symmetry that explicitly forbids the tadpole term, the non-renormalization theorem will prevent it from being generated radiatively until SUSY is broken. We will assume that the tadpole term is either absent or simply too small to play any role in the dynamics of the model. Finally, while the $S^3$ term is required in the usual NMSSM in order to stabilize the potential in the $S$ direction, our potential is stabilized by the explicit mass term, $\mu_s$. Because it is no longer required, and because its effects will tend to be small anyway, we will take $\kappa$ to be effectively zero in the analysis that follows. Therefore the superpotential of interest here is given by 
\begin{equation}\label{W}
W=(\mu+\lambda {S}) H_{u}{H_{d}}+\frac{1}{2}\mu_{s}{S}^{2}.
\end{equation}
We will refer to this very simple singlet extension of the MSSM as the S-MSSM hereafter. Philosophically, the S-MSSM is an attempt to {\it barely}\/ extend the MSSM; not only is the particle content minimally extended, but the structure of the vacuum will be nearly identical to that of the MSSM. In this sense, our philosophy is similar to that in Ref.~\cite{dst}, and indeed those authors briefly considered a model like the S-MSSM, though without examining its implications in any detail.

The Lagrangian for the scalar fields is given by the familiar $F$- and $D$-terms
as well as soft-breaking operators. These are given by the usual MSSM soft-breaking terms, including the bilinear $B_\mu H_u H_d$, plus contributions from the gauge singlet. The terms relevant to the Higgs potential are:
\begin{equation}\label{eq:three}
V_{\rm soft}=m^{2}_{s}|S|^{2}+m_{H_u}^{2}|H_{u}|^{2}+m_{H_d}^{2}|H_{d}|^{2}+(B_\mu H_{u}H_{d}+B_s S^2 + \lambda A_{\lambda}SH_{u}H_{d}+h.c.)+\cdots
\end{equation}
The scalar potential in the Higgs-singlet sector is thus given by
\begin{eqnarray}\label{eq:four}
V&=&(m^{2}_{H_u}+|\mu+\lambda S|^2)|H_{u}|^{2}+(m^{2}_{H_d}+|\mu+\lambda S|^2)|H_{d}|^{2}+(m_{s}^{2}+\mu_{s}^{2})|S|^{2} \nonumber \\ 
&&+\,\left[B_s S^2+\left(\lambda\mu_{s}S^{\dagger}+B_\mu+\lambda A_{\lambda} S\right)H_{u}H_{d}+h.c.\right] + \lambda^{2}\left|H_{u}H_{d}\right|^2 \nonumber \\
&&+\,\frac{1}{8}(g^{2}+g'^{2})\left(|H_{u}|^{2}-|H_{d}|^{2}\right)^{2} +\frac{1}{2}g^{2}|H_{u}^{\dagger}H_{d}|^{2}.
\end{eqnarray}
On minimizing this potential, one finds three conditions similar in form both to the MSSM and NMSSM:
\begin{eqnarray}
\frac12 \mz^2 &=& \frac{m^2_{H_d} - m^2_{H_u}\tan^2\beta}{\tan^2\beta-1} -\mueff^2, \\
\sin 2\beta &=& \frac{2\Bmueff}{m^2_{H_u} + m^2_{H_d} + 2\mueff^2 + \lambda^2 v^2},
\end{eqnarray}
and
\beq
v_s=\frac{\lambda v^2}{2}\, \frac{(\mu_{s}+A_{\lambda})
\sin 2\beta-2\mu}{\lambda^{2}v^{2}+\mu_{s}^{2}+m_{s}^{2}+2B_{s}} \simeq \frac{\lambda v^2}{2\mu_s}\sin 2\beta, \label{vs}
\eeq
where $v_s=\vev{S}$ and $v_{u,d}=\vev{H_{u,d}}$, with $v=(v_u^2+v_d^2)^{1/2}$ as previously defined; and
\begin{eqnarray}
\mueff &=& \mu + \lambda v_s, \\
\Bmueff &=& B_\mu + \lambda v_s (\mu_s+A_\lambda ).
\end{eqnarray}
The second equality in Eq.~(\ref{vs}) holds when $\mu_s$ is taken large compared to the other mass parameters, and it demonstrates that
while the S-MSSM is superficially very similar to the NMSSM (it has the same particle content), its vacuum structure can be quite dissimilar. In particular, the S-MSSM possesses a limit, $\mu_s\to\infty$ and $v_s\to 0$, in which the singlet can be integrated out supersymmetrically, leaving the MSSM as the low-energy model. As we approach this limit, the vacuum structure of the S-MSSM resembles that of the MSSM in that the minimum of the potential is at $v_{u,d}\sim 100\gev$ but $v_s\sim\,$few GeV. In fact, even when $\mu_s$ is not particularly large compared to the other soft masses, the vev of $S$ is highly suppressed. In contrast, in the NMSSM, the singlet can only be integrated out non-supersymmetrically, by taking $|m_s^2|$ large, leaving behind the MSSM plus the fermionic component of the singlet superfield. Further, the vev of $S$ must be large in order to generate an effective $\mu$-term, $\mueff = \lambda v_s$, consistent with chargino and neutralino mass bounds. But in order to avoid doublet vevs that are too small, the singlet must not be so large as to over-stabilize the Higgs potential, constraining $v_s$ both from below and above.

The physical spectrum of the Higgs sector includes a single charged Higgs boson ($H^\pm$), three neutral scalars ($h^0$, $H^0_{1,2}$), and two neutral pseudoscalars ($A^0_{1,2}$). In the discussion presented here, we will assume that $\mu_s$ is the largest mass scale in the Higgs sector; were $\mu_s$ to be small compared to $\mu$ and the soft masses, one would find substantial mixing between the singlets and the Higgs doublets, which would reduce the mass of the lightest Higgs. But for large $\mu_s$ we can find simple formulae for the Higgs masses as an expansion in inverse powers of $\mu_s$. The ``heavy" neutral Higgs bosons have masses:
\begin{eqnarray}
m_{A^{0}_{1}}^{2}&\simeq&\frac{2B_\mu}{\sin2\beta}+\frac{4\lambda^{2}A_{\lambda}v^{2}}{\mu_{s}}-\frac{2\lambda^{2}\mu v^{2}}{\mu_{s}\sin2\beta} \\
m^2_{A^{0}_{2},H^{0}_{2}}&\simeq&\mu_{s}^{2}+2\lambda^{2}v^{2}+m_{s}^{2}\mp 2B_{s}
\end{eqnarray}
where the approximation holds for $\mu_s$ much greater than all the other mass scales.
Meanwhile the scalar Higgs bosons, $h^0$ and $H^0$, receives masses:
\begin{equation}
m^{2}_{h^{0},H^0_1}\simeq \left.m^{2}_{h^0,H^0_1}\right|_{\mbox{\scriptsize MSSM}}+\frac{2\lambda^{2}v^{2}}{\mu_{s}}\left(\mu\sin2\beta-A_{\lambda}\mp\Delta^{2}\right)
\end{equation}
where the first term is the mass of the lightest Higgs scalar as calculated in the MSSM (replacing $m_{A^0}\to m_{A^0_1}$), and 
\begin{equation}
\Delta^{2}=\frac{A_{\lambda}(m^{2}_{Z}-m^{2}_{A^{0}_1})\cos^{2}2\beta-\mu(m^{2}_{A^{0}_1}+m^{2}_{Z})\sin2\beta}{\sqrt{(m^{2}_{A^{0}_1}+m^{2}_{Z})^{2}-4m^{2}_{A^{0}_1}m^{2}_{Z}\cos^{2}2\beta}}.
\end{equation}
The expression for the $h^0$ mass in the S-MSSM is most easily understood in the Higgs decoupling limit, $m_{A^0_{1,2}}\to \infty$, in which it is also maximized:
\begin{equation} \label{mhsmssm}
m^2_{h^0} \simeq \mz^2\cos^2 2\beta + \frac{2\lambda^2 v^2}{\mu_s}\left(2\mu\sin 2\beta - A_\lambda\sin^2 2\beta\right) . \label{hmass}
\end{equation}
We see in Eq.~(\ref{hmass}) the expected behavior that as $\mu_s\to\infty$, the usual bound on the lightest Higgs mass from the MSSM is recovered. (Note that we use the full, exact expressions for the masses in the numerical analysis below.)

It is natural to compare this result to the expression for the lightest scalar mass in the NMSSM, Eq~(\ref{nmssmresult}); one finds several key differences. First, the expression in Eq.~(\ref{nmssmresult}) is only an upper bound. In fact, it is an eigenvalue of the $2\times2$ submatrix corresponding to the $\{\mbox{Re}\,H_u^0,\mbox{Re}\,H_d^0\}$ entries in the full $3\times3$ scalar mass matrix. Mixing of the light Higgs state with the scalar component of $S$ will suppress its mass, and so one usually tries to tune the mixing to be very small in the NMSSM. This is done by tuning the off-diagonal elements in the scalar mass matrix: 
\beq 
A_\lambda\simeq \frac{2\mu}{\sin 2\beta} - 2\kappa v_s. \quad\quad\mbox{(NMSSM)}
\label{nmtune}
\eeq 
In fact one finds that the mass of the lightest Higgs in the NMSSM falls very rapidly as one moves away from this point. On the other hand, Eq.~(\ref{hmass}) holds in the S-MSSM in a wide range of parameter space, as it already includes the effects of mixing at leading order in $1/\mu_s$. 

Second the NMSSM contribution to $m^2_{h^0}$ falls as $\sin^2 2\beta$, so that the effect of the singlet drops rapidly as one increases $\tan\beta$ above one. This is unfortunate for the NMSSM, because it means that the model only provides large contributions to the lightest Higgs mass precisely when the tree-level MSSM contribution is minimized. Though the situation is similar in the S-MSSM, here there are two terms which contribution to $m^2_{h^0}$, one that falls as $\sin^2 2\beta$ as in the NMSSM, but another that falls only as $\sin 2\beta$, which allows for new contributions at intermediate values of $\tan\beta$, and thus potentially larger masses.

We now analyze the S-MSSM in order to test whether we can obtain light Higgs masses above the LEP bound of $114\gev$ while keeping our top squark spectrum below $1\tev$. 
Since we have not embedded this low-energy model into a full-blown model of SUSY breaking, we cannot speak with precision about the fine tuning inherent in our calculations, but the requirement that the top squarks (and all other SUSY mass parameters) fall below a TeV is an oft-used substitute for a full fine-tuning analysis. 

It is not at all obvious that the S-MSSM can generate such large Higgs masses. In the NMSSM, Ref.~\cite{ellhug} was able to push the Higgs mass all the way to $141\gev$ with $m_{\tilde t} = 1\tev$ and $A_t=2.5\tev$, as long as the condition in Eq.~(\ref{nmtune}) holds. But the S-MSSM more closely resembles the MSSM and so the vacuum structure, and corresponding scalar mass matrix, bears little resemblance to the NMSSM case. In fact, we will find that the S-MSSM works equally well at generating large Higgs masses.

Our calculations of the Higgs masses are done using the full 1-loop effective potential, including contributions from the singlet superfield. Since it is well known that the 2-loop contributions can severely suppress the light Higgs mass, we include the leading 2-loop contributions (dominated by gluinos) as calculated in FeynHiggs~\cite{feynhiggs}; these negative contributions are added in quadrature to the result of our 1-loop effective potential analysis. Though the leading-order 2-loop effects are the same in the MSSM and S-MSSM, sub-leading terms may differ, potentially introducing a small error into our analysis.

In order to maximize the light Higgs mass, we take $m_{A^0_1}$ to be large, effectively decoupling one of the Higgs doublets from the electroweak symmetry-breaking sector. We also maximize the coupling $\lambda$. However, because one of the advantages of singlet extensions of the MSSM is the preservation of gauge coupling unification as in the MSSM, we insist that $\lambda$ remain perturbative at all scales up to $2\times 10^{16}\gev$. This places an upper bound on $\lambda$ which varies with $\tan\beta$: $\lambda < 0.4$ at $\tan\beta=1.5$, but quickly rises to $\lambda <0.6$ by $\tan\beta=2$ and to $\lambda<0.7$ for $\tan\beta\geq3$. This requirement is also commonly enforced in analyses of the NMSSM~\cite{quiesp,ellhug}. In our case the renormalization group equation for $\lambda$ is identical to that of the NMSSM, except that we can set the coupling $\kappa$ in Eq.~(\ref{Wfull}) to zero, increasing our allowed $\lambda$ very slightly.

\begin{figure}[htb]
\begin{center}
\includegraphics[width=0.8\linewidth]{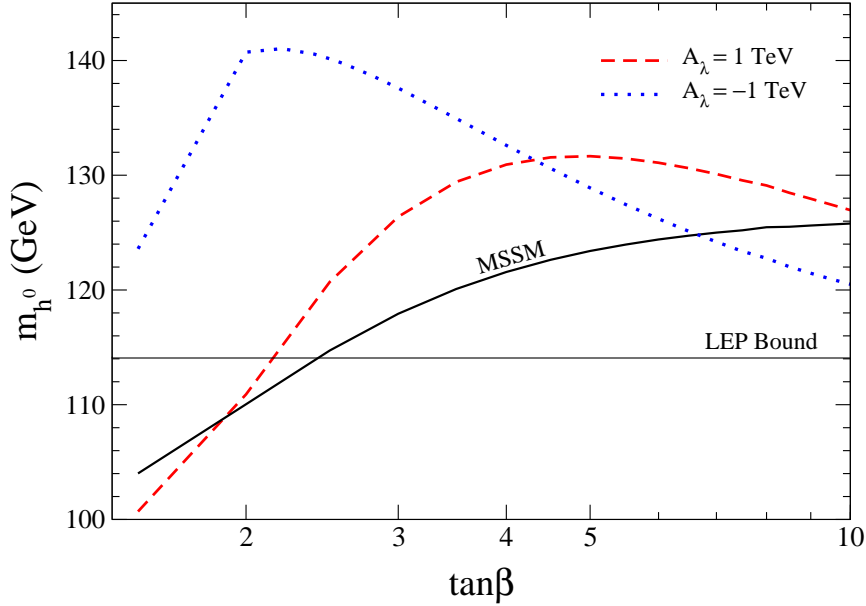}
\caption{Lightest neutral Higgs mass as a function of $\tan\beta$ in the MSSM and S-MSSM. The red dashed/blue dotted curves were obtained using $\mu_s=m_{\tilde t}=M_{\tilde g}=1\tev$ and $A_{\lambda}=\pm1\tev$ in the S-MSSM. The solid black curve represents the MSSM. See text for additional parameters used in the figure.}\label{fig1}
\end{center}
\end{figure}

Our main result is shown in Figure~\ref{fig1}, where we have plotted the lightest Higgs mass as a function of $\tan\beta$ for a typical set of S-MSSM input parameters. In particular, for this figure we have chosen $\mu_s=2\tev$, $m_{\tilde t}=M_{\tilde g}=1\tev$, $\mu=500\gev$, and $A_\lambda= \pm 1\tev$ on the red dashed/blue dotted lines respectively. The value of $B_s$ has been chosen to be $-(100\gev)^2$ for the figure, but makes almost no difference in the analysis; $B_\mu$ is calculated as an output of the minimization procedure. In order to maximize the Higgs mass, we are working in what is essentially the ``maximal mixing" case of the MSSM in which $A_t=\sqrt{6}\, m_{\tilde t}\simeq 2.5\tev$. The black dashed line represents the upper bound on the MSSM Higgs mass for the same set of input parameters.

We note two results from the figure. First, for $A_\lambda=-1\tev$ we are able to get light Higgs masses above $140\gev$ at $\tan\beta\simeq 2.2$, and able to get $m_{h^0}>130\gev$ over a wide range of $\tan\beta\lsim 5$. For $A_\lambda=+1\tev$ we can also get $m_{h^0}>130\gev$, but for larger $\tan\beta$. Thus our results are rather robust against varying $\tan\beta$. Second, we note that different signs of $A_\lambda$ maximize the Higgs mass for different $\tan\beta$. This behavior is not apparent from Eq.~(\ref{hmass}), but arises at next order in the $1/\mu_s$ expansion.

\begin{figure}[htb]
\begin{center}
\includegraphics[width=0.8\linewidth]{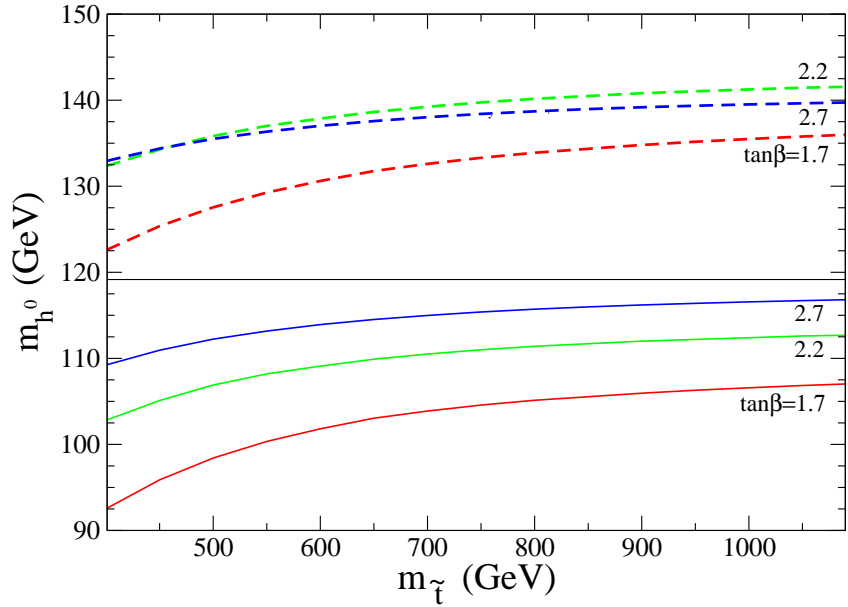}
\caption{Lightest neutral Higgs mass as a function of $m_{\tilde{t}}$ in the MSSM (solid) and S-MSSM (dashed) for three values of $\tan\beta$ and assuming maximal mixing. See text for additional parameters used in the figure.}\label{fig2}
\end{center}
\end{figure}

The SUSY-breaking masses in Figure~\ref{fig1} are still rather large, so it is natural to ask how dependent our Higgs mass is on the stop-loop-induced contributions. In Figure~\ref{fig2} we show the dependence of the lightest Higgs mass with respect to the SUSY-breaking masses (upper dashed lines). We set $m_{\tilde t}=M_{\tilde g}$ and vary them between $400\gev$ and $1.1\tev$, all the while setting $A_t=\sqrt{6} m_{\tilde t}$, $\mu=500\gev$ and $\mu_s=2\tev$, for three different values of $\tan\beta$. As can be seen, even for a very light spectrum ($m_{\tilde t}\approx 400\gev$) the mass of the lightest Higgs mass is well above the LEP bound whereas for the same choice of parameters the MSSM will predict a light Higgs which is ruled out by experimental searches (lower solid lines).

It is natural to ask how the S-MSSM might be distinguished from the MSSM or the NMSSM at the LHC. For the range of parameters considered here, where $\mu_s$ is relatively large compared to the other masses, the singlet effectively decouples, with only small mixings into the Higgs bosons and neutralinos. Neither does the $S$-field play any significant role in the model's dark matter candidate. Thus the dominant signature for the S-MSSM (and for the NMSSM in much of its parameter space) is a Higgs mass measured to be larger than allowed by the observed top squark mass spectrum. Otherwise the model mimics the MSSM.

In conclusion, by marrying the Next-to-Minimal SUSY Standard Model to explicit, supersymmetric mass terms, we sacrifice the ability to solve the $\mu$-problem of the MSSM but gain the ability to alleviate, or even solve, the little hierarchy problem. The S-MSSM allows light Higgs masses well above $114\gev$ even for top squark masses well below the TeV scale, and does so without requiring any cancellations among various model parameters. Of course, the model presented here should only be considered the infrared limit of a more complete model in which the SUSY-breaking masses are generated~\cite{us} and perhaps also the SUSY-preserving masses. In such a model the issue of fine tuning could be tackled in a more complete fashion. But from a strictly low-energy point of view, this model satisfies all experimental constraints, including the LEP Higgs bound, without resorting to large loop corrections or fine-tuned cancellations among parameters. Thus it would be a natural candidate for explaining the observation of SUSY with light stops at the LHC. 

\section*{Acknowledgments}
A.D. and C.K. want to thank the Aspen Center for Physics where this project was begun, and Lisa Everett for helpful discussions. This work was partly supported by the National Science Foundation under grant PHY-0905383-ARRA.

\end{document}